\documentclass[twocolumn,showpacs,preprintnumbers,amsmath,amssymb]{revtex4}

\usepackage{graphicx}
\usepackage{dcolumn}
\usepackage{bm}
\usepackage{rotating}

\begin{document}

\title{Precision spectroscopy of the hydrogen molecular ion HD$^+$}

\author{Zhen-Xiang Zhong$^{1}$, Pei-Pei Zhang$^{1,2}$, Zong-Chao Yan$^{1,3}$ and Ting-Yun Shi$^{1}$}
\affiliation{$^{1}$Division of Theoretical and Interdisciplinary Research,
State Key Laboratory of Magnetic Resonance and Atomic and Molecular Physics,
Wuhan Institute of Physics and Mathematics,
Chinese Academy of Sciences, Wuhan 430071, China}
\affiliation{$^{2}$Graduate School of the Chinese Academy of
Sciences, Beijing 100049, China}
\affiliation{$^{3}$Department of Physics, University of New Brunswick,
Fredericton, New Brunswick, Canada E3B 5A3}
\date{\today}

\begin{abstract}
Expectation values of the Breit operators and the $Q$ terms are calculated for HD$^+$ with the vibrational number $v=0\!-\!4$ and the total angular momentum $L=0\!-\!4$. Relativistic and radiative corrections to some ro-vibrational transition frequencies are determined. Numerical uncertainty in $R_{\infty}\alpha^2$ order correction is reduced to sub kHz or smaller. Our work provides an independent verification of Korobov's calculations [Phys. Rev. A {\bf74}, 052506 (2006); {\bf77}, 022509 (2008)].
\end{abstract}

\maketitle


\section{Introduction}

Hydrogen molecular ions, such as H$_2^+$ and HD$^+$, can be used \cite{Hil00,Sch05} to deduce an improved value of the proton-electron mass ratio by comparing experimental and theoretical spectroscopic data. To this end, several experiments have been setup for measuring high precision transition frequencies in HD$^+$~\cite{Koe07,Koe11} and ${\rm H}_2^+$~\cite{Kar08}.
For HD$^+$ in particular, the $(v,L)=(4,3)\rightarrow (0,2)$ transition frequency has been measured at the 2 ppb level~\cite{Koe07}.
In order to reduce the current uncertainty of $4.1\times10^{-10}$~\cite{CO10} in the proton-electron mass ratio, both experiment and theory should reach a precision of sub-kHz or better.

For light systems, nonrelativistic QED (NRQED) \cite{Cas86,Nio97} approach is used to expand the energy in powers of
the fine structure constant $\alpha$. Nonrelativistic energies of ${\rm H}_2^+$ and ${\rm HD}^+$ have been
variationally calculated at the precision of $10^{-15}$~\cite{Kar05,Sch05} for a wide range of vibrational states
and at the precision of $10^{-30}$~\cite{Kor00,Bai02,Yan03,Cas04,Li07} for some low vibrational states.
Relativistic and radiative corrections of order $R_{\infty}\alpha^2$, $R_{\infty}\alpha^3$, $R_{\infty}\alpha^4$, and $R_{\infty}\alpha^5$ have been systematically calculated~\cite{Kor06,Kor08} for ${\rm H}_2^+$ and ${\rm HD}^+$ ro-vibrational states ($v=0\!-\!4,L=0\!-\!4$). The leading order $R_{\infty}\alpha^2$ relativistic corrections are now available with very high precision for the ${\rm H}_2^+$ vibrational states ($0,0$), ($1,0$), and ($0,1$)~\cite{Zho09}. Recently, improved values of $R_{\infty}\alpha^3$ order corrections have been achieved by an intensive calculation of the Bethe logarithm term~\cite{Kor12}.

The purpose of this Brief Report is to present our independent calculations of the Breit operators and the $Q$ terms for the HD$^+$ ro-vibrational states ($v=0\!-\!4,L=0\!-\!4$). Some important ro-vibrational transition frequencies are determined, which provides a verification of Korobov's theoretical work. In addition, the numerical results presented here might serve as a benchmark for other theoretical methods.
We use atomic units ($\hbar=e=m_e=1$), unless otherwise stated. The fundamental physical constants involved  are taken from the 2010 CODATA recommended values~\cite{CO10}.

\section{theory}

Consider the hydrogen molecular ion HD$^+$. After separating the center of mass coordinates for the system, the eigenvalue problem for the nonrelativistic Hamiltonian $H_0$ becomes~\cite{Li07}
\begin{align}{\label{eq:schr}}
H_0&\Psi = E_0 \Psi\,,\\
H_0&=\lambda_1\nabla_{r_1}^2+\lambda_2\nabla_{r_2}^2+\lambda_{12}\nabla_{r_1}\cdot\nabla_{r_2}+V\,,
\end{align}
 where ${\bf r}_1$ and ${\bf r}_2$ represent respectively the position vectors of the electron and proton, relative to the deuteron situated at the origin, $\lambda_1 = -(1+m_d)/(2m_d)$, $\lambda_2 = -(1/m_d+1/m_p)/2$, $\lambda_{12} = -1/m_d$,  $V=-1/r_1+1/r_2-1/r_{12}$ is the Coulomb interaction, and $\mathbf{r}_{12}=\mathbf{r}_1-\mathbf{r}_2$. The energy eigenvalue problem for $H_0$ is solved variationally using the basis set in Hylleraas coordinates
\begin{equation}\label{eq:hyl}
\phi_{ijk}({\bf r}_1,{\bf r}_2)=r_1^{i} r_2^{j} r_{12}^{k}
e^{-\alpha r_1 -\beta r_2}\mathcal{Y}^{LM}_{l_1 l_2}({\bf r}_1,{\bf
r}_2)\,,
\end{equation}
where $\mathcal{Y}^{LM}_{l_1 l_2}({\bf r}_1,{\bf r}_2)$ is the vector coupled product of spherical harmonics for the electron and the proton to form a common eigenvector of $L^2$ and $L_z$. More details on the construction of basis set for HD$^+$ may be found in~\cite{Yan03,Li07}. It should be pointed out that this basis set differs from the one used by Korobov~\cite{Kor06,Kor08}. The basic type of integrals required in the calculation of matrix elements can be evaluated analytically~\cite{Yan96} using Perkins' expansion for $r_{12}^k$. The procedure for handling singular integrals that appear in the evaluation of Breit operators can be found
in~\cite{Yan94}.

The leading-order relativistic corrections due to the Breit operators are well established, which can be found in Refs.~\cite{Bet77,Ber82,Eid01}.
The complete spin-independent part of order $R_{\infty}\alpha^2$ term is
\begin{align}\label{eq:aa2}
  &E^{(2)}=\alpha^2 \langle H_{\rm Breit}\rangle\,,
\end{align}
where
{\small
\begin{eqnarray}\label{eq:a2}
  H_{\rm Breit}&=&-\frac{1}{8}{\nabla_{r_1}^4}-\frac{1}{8m_p^3}{\nabla_{r_2}^4}
  -\frac{1}{8m_d^3}(\nabla_{r_1}+\nabla_{r_2})^4\nonumber\\
  &+&\frac{\pi}{2m_p^2}\delta(\mathbf{r}_{12})
  +\frac{\pi}{2}\big[\delta(\mathbf{r}_1)+\delta(\mathbf{r}_{12})\big]\nonumber\\
  &-&R_{de}+R_{dp}-R_{pe}\,,
\end{eqnarray}
}
where
{\small
\begin{eqnarray*}
  {R}_{de}&=&\frac{-1}{2m_d}\left(\frac{\nabla_{r_1}\cdot(\nabla_{r_1}+\nabla_{r_2})}{r_1}
  +\frac{\mathbf{r}_1\mathbf{r}_1:(\nabla_{r_1}+\nabla_{r_2})\nabla_{r_1}}{r_1^3}\right)\\
  {R}_{dp}&=&\frac{-1}{2m_dm_p}\left(\frac{\nabla_{r_2}\cdot(\nabla_{r_1}+\nabla_{r_2})}{r_2}
  +\frac{\mathbf{r}_2\mathbf{r}_2:(\nabla_{r_1}+\nabla_{r_2})\nabla_{r_2}}{r_2^3}\right)\\
  {R}_{pe}&=&\frac{1}{2m_p}\left(\frac{\nabla_{r_1}\cdot\nabla_{r_2}}{r_{12}}
  +\frac{\mathbf{r}_{12}\mathbf{r}_{12}:\nabla_{r_1}\nabla_{r_2}}{r_{12}^3}\right)\,.
\end{eqnarray*}
}In the above, the Darwin term $\pi/(2m_p^2)\delta(\mathbf{r}_{12})$ is the
nuclear spin dependent recoil correction for the spin-$\frac{1}{2}$ particle, such as proton.
This term vanishes in case of spin-0 or spin-1 nucleus, such as the $^4$He nucleus or deuteron~\cite{pachu_jpb95}. It is noted that
$\delta({\bf r}_2)$ is virtually zero due to the molecular nature of the system.

Furthermore, the spin-independent radiative correction of order $R_{\infty}\alpha^3$
may be expressed as~\cite{Kor04,Pac98,Yel01}:
\begin{align}\label{eq:a3}
  &E^{(3)}=\alpha^3\bigg\{\frac{4}{3}\left[-\ln\alpha^2-\beta(v,L)
  +\frac{19}{30}\right]\langle\delta(\mathbf{r}_1)+\delta(\mathbf{r}_{12})\rangle\nonumber\\
  &+\frac{2}{3}\left[-\ln\alpha-4\beta(v,L)+\frac{31}{3}\right]\left[\frac{\langle\delta(\mathbf{r}_1)\rangle}{m_d}
  +\frac{\langle\delta(\mathbf{r}_{12})\rangle}{m_p}\right]\nonumber\\
  &-\frac{14}{3}\left[\frac{Q({\bf r}_1)}{m_d}+\frac{Q({\bf r}_{12})}{m_p}\right]\bigg\}\,,
\end{align}
where $\beta(v,L)$ is the Bethe logarithm, $Q({\bf r}_1)$ and $Q({\bf r}_{12})$ are
the $Q$ terms introduced by Araki and Sucher~\cite{Qterm},
\begin{equation}
  Q({\bf r})=\lim_{\rho\rightarrow0}\left\langle\frac{\Theta(r-\rho)}{4\pi{r}^3}+(\ln\rho+\gamma_E)\delta(\mathbf{r})\right\rangle\,,
\end{equation}
and $\gamma_E$ is the Euler gamma constant.

For higher order corrections, such as orders $R_{\infty}\alpha^4$ and $R_{\infty}\alpha^5$, we follow the work of Ref.~\cite{Kor08}.
Thus, $R_{\infty}\alpha^4$ order non-recoil relativistic and radiative corrections may be express as follows \cite{Kor08}:
\begin{align}\label{eq:a4}
  &E^{(4)}=\alpha^4\bigg\{\frac{1}{\pi}\left[-\frac{2179}{648}+\frac{3523}{864}\pi^2-\frac{1}{2}\pi^2\ln2-\frac{9}{4}\zeta(3)\right]\nonumber\\
  &\times\langle\delta(\mathbf{r}_1)+\delta(\mathbf{r}_{12})\rangle+E_{\rm{rel}}^{(4)}\bigg\}\,,
\end{align}
where $E_{\rm{rel}}^{(4)}$ is the $R_{\infty}\alpha^4$ order relativistic correction.

Since the electron is almost bounded in the ground state of hydrogen molecular ion, its wave function can be approximately expressed as a linear combination of two hydrogen-like wave functions $\psi_e(\mathbf{r}_e)=C[\psi_{1s}(\mathbf{r}_1)+\psi_{1s}(\mathbf{r}_{12})]$. Therefore, the most important $R_{\infty}\alpha^5$ order correction can be estimated using this approximate wave function~\cite{Kor08,Eid01}
\begin{align}\label{eq:a5}
  &E^{(5)}=\alpha^5\bigg[-\ln^2\frac{1}{(\alpha)^2}+A_{61}\ln\frac{1}{(\alpha)^2}+A_{60}+\frac{B_{50}}{\pi}\bigg]\nonumber\\
  &\times\langle\delta(\mathbf{r}_1)+\delta(\mathbf{r}_{12})\rangle\,,
\end{align}
where the constants $A_{61}$, $A_{60}$, and $B_{50}$ are taken to be the constants of the $1s$
state of the atomic hydrogen, {\it i.e.}, $A_{61}=5.419\cdots$ \cite{Lay60}, $A_{60}=-30.924\cdots$ \cite{Pau93}, and $B_{50}=-21.556\cdots$ \cite{Pau94}.

In addition to the relativistic and radiative corrections, one also needs to consider the contribution from the finite nuclear charge
distribution. The leading-order correction is
\begin{equation}\label{eq:nuc}
  E_{\rm nuc}=\frac{2\pi}{3}\left[\left(\frac{R_d}{a_0}\right)^2\langle\delta(\mathbf{r}_1)\rangle
  +\left(\frac{R_p}{a_0}\right)^2\langle\delta(\mathbf{r}_{12})\rangle\right]\,,
\end{equation}
where $R_p=0.8775(51)$ fm and $R_d=2.1424(21)$ fm~\cite{CO10} are the root-mean-square charge radii of proton and deuteron respectively.

\section{results}

\begin{table}
  \caption{Convergence study of the expectation values of $\delta(\mathbf{r}_{1})$ and $\delta(\mathbf{r}_{12})$ for the HD$^+$ ($v=4,L=4$) state, where $N$ is number of basis functions.}
  \label{tb:hde:convergent}
  \begin{tabular}{c@{\hspace{5mm}}l@{\hspace{5mm}}l}
  \hline\hline
    $N$ & \multicolumn{1}{c}{$\delta(\mathbf{r}_{1})$} & \multicolumn{1}{c}{$\delta(\mathbf{r}_{12})$} \\\hline
   2789 & 0.188\ 729\ 941\ 487 & 0.188\ 382\ 948\ 5531\\
   3415 & 0.188\ 729\ 516\ 647 & 0.188\ 382\ 898\ 6629\\
   4130 & 0.188\ 729\ 943\ 378 & 0.188\ 382\ 886\ 4800\\
   4940 & 0.188\ 729\ 938\ 954 & 0.188\ 382\ 884\ 0699\\
   5851 & 0.188\ 729\ 935\ 696 & 0.188\ 382\ 876\ 7837\\
   6859 & 0.188\ 729\ 935\ 329 & 0.188\ 382\ 876\ 7442\\
   8000 & 0.188\ 729\ 935\ 332 & 0.188\ 382\ 876\ 7368\\
   9250 & 0.188\ 729\ 935\ 329 & 0.188\ 382\ 876\ 7376\\
  Extrp. & 0.188\ 729\ 935\ 323(7) & 0.188\ 382\ 876\ 7375(1)\\
  \hline\hline
  \end{tabular}
\end{table}

 With the Hylleraas-type basis set of Eq. (\ref{eq:hyl}), the wave functions along with the corresponding nonrelativistic energies are obtained by solving Eq. (\ref{eq:schr}) variationally. Then the expectation values of the Breit operators can be evaluated.
 In particular, the global operator method~\cite{Dra81} is applied to the evaluations of $\nabla_{r_1}^4$, $\nabla_{r_2}^4$, $(\nabla_{r_1}+\nabla_{r_2})^4$, $\delta({\bf{r}}_1)$, and  $\delta({\bf{r}}_{12})$. As an example, Table~\ref{tb:hde:convergent} shows a convergence study for
 the expectation values of $\delta({\bf{r}}_{1})$ and $\delta({\bf{r}}_{12})$.  One can see that an accuracy of about 11-12 significant figures is achieved for the most difficult state of ($4,4$), where the nonrelativistic energy is calculated only to 16 digits. For ro-vibrational states ($v=0\!-\!4,L=0\!-\!4$), numerical results of the Breit operators are presented in
 Tables~\ref{tb:hde:breit1} and \ref{tb:hde:breit2}, together with a comparison with Korobov's values~\cite{Kor06}.
 Results for the $Q$ terms are listed in Tables~\ref{tb:hde:qtm1} and \ref{tb:hde:qtm2}.
 All expectation values are in good agreement with Korobov's values, although our results are more precise.
 The largest size of basis set used here is about 9000.

\begin{table*}
\caption{Expectation values of $\delta({\bf{r}}_{1})$, $\delta({\bf{r}}_{12})$, $\nabla_{r_1}^4$, $\nabla_{r_2}^4$, and $(\nabla_{r_1}+\nabla_{r_2})^4$ for HD$^+$ with $v=0-4$ and $L=0-4$. Korobov's results~\cite{Kor06} are listed in the second entry of each ro-vibrational state. }
\label{tb:hde:breit1}
\begin{tabular}{clllll}
\hline\hline
$(v,L)$ & \multicolumn{1}{c}{$\delta({\bf{r}}_{1})$} & \multicolumn{1}{c}{$\delta({\bf{r}}_{12})$} & \multicolumn{1}{c}{$\nabla_{r_1}^4$} & \multicolumn{1}{c}{$\nabla_{r_2}^4$} & \multicolumn{1}{c}{$(\nabla_{r_1}+\nabla_{r_2})^4$} \\\hline
$(0,0)$     & 0.2073481417802970666(2)  & 0.2070425994774719004(2)  & 6.3001999476785933(6) & 104.371713686093(1)   & 104.443848901053(2) \\
            & 0.207348142               & 0.207042599               & 6.30019995            & 104.372               & 104.444 \\
$(1,0)$     & 0.20260117861162289(2)    & 0.20228886473978993(2)    & 6.15902235229201(4)   & 449.45675946471(2)    & 449.7391462670(1)   \\
            & 0.202601179               & 0.202288865               & 6.15902236            & 449.457               & 449.739 \\
$(2,0)$     & 0.1981667951256018414(8)  & 0.19784583746893604(2)    & 6.02761432267228(1)   & 1042.81683129264(3)   & 1043.44331653088(5) \\
            & 0.198166795               & 0.197845838               & 6.02761433            & 1042.82               & 1043.443 \\
$(3,0)$     & 0.1940278413804545(3)     & 0.1936960139249636(3)     & 5.9054534324885(7)    & 1812.997565364(1)     & 1814.063236354(2)   \\
            & 0.194027841               & 0.193696014               & 5.90545344            & 1813.00               & 1814.06 \\
$(4,0)$     & 0.19016909089076(3)       & 0.189823702852585(7)      & 5.7920773795339(5)    & 2697.669274(1)        & 2699.2354157(9)     \\
            & 0.190169092               & 0.189823704               & 5.79207742            & 2697.67               & 2699.24 \\
$(0,1)$     & 0.207163241677423572(7)   & 0.20685769957275256(1)    & 6.294450746055923(5)  & 110.38493981778(2)    & 110.4613313632(1) \\
            & 0.207163242               & 0.206857700               & 6.29445075            & 110.385               & 110.461 \\
$(1,1)$     & 0.20242655716340313(3)    & 0.20211421220190804(1)    & 6.15359974575108(2)   & 464.36193814804(1)    & 464.6533872074(1) \\
            & 0.202426557               & 0.202114213               & 6.15359977            & 464.362               & 464.653 \\
$(2,1)$     & 0.198001981206342(1)      & 0.1976809559785677(9)     & 6.0225036774882(2)    & 1064.624132727(1)     & 1065.263406622(4) \\
            & 0.198001983               & 0.197680957               & 6.02250376            & 1064.62               & 1065.26 \\
$(3,1)$     & 0.193872416785920(1)      & 0.1935404783766321(6)     & 5.9006418093939(4)    & 1839.951687858(5)     & 1841.032921943(3) \\
            & 0.193872418               & 0.193540480               & 5.90064189            & 1839.95               & 1841.03 \\
$(4,1)$     & 0.1900226872284(1)        & 0.18967713560504(1)       & 5.787553428565(6)     & 2728.2144158(1)       & 2729.7980517(2) \\
            & 0.190022692               & 0.189677140               & 5.78755367            & 2728.21               & 2729.80 \\
$(0,2)$     & 0.2067954443349440(4)     & 0.20648990097608551(7)    & 6.2830163299151(5)    & 123.79001412764(6)    & 123.8756587418(6)\\
            & 0.206795445               & 0.206489902               & 6.28301636            & 123.790               & 123.876 \\
$(1,2)$     & 0.2020792303568918(5)     & 0.2017668215015289(1)     & 6.142815739370(1)     & 495.3679016186(8)     & 495.678116390(6)\\
            & 0.202079231               & 0.201766822               & 6.14281577            & 495.368               & 495.678 \\
$(2,2)$     & 0.1976741873189976(5)     & 0.197353025118044(5)      & 6.012340962862(5)     & 1109.271531087(7)     & 1109.93693530(3) \\
            & 0.197674189               & 0.197353027               & 6.01234104            & 1109.27               & 1109.94 \\
$(3,2)$     & 0.193563323743184(3)      & 0.193231161405204(1)      & 5.89107464757(3)      & 1894.7472155(5)       & 1895.8600500(8)\\
            & 0.193563325               & 0.193231163               & 5.89107470            & 1894.75               & 1895.86 \\
$(4,2)$     & 0.18973156232654(8)       & 0.18938568121580(3)       & 5.77855925079(2)      & 2790.0612310(3)       & 2791.6802590(4)\\
            & 0.189731566               & 0.189385688               & 5.77855952            & 2790.06               & 2791.68 \\
$(0,3)$     & 0.2062486970164612(4)     & 0.20594314768281(1)       & 6.266022789768(7)     & 147.25876381(2)       & 147.36009295(4)\\
            & 0.206248697               & 0.205943148               & 6.26602281            & 147.259               & 147.360 \\
$(1,3)$     & 0.2015629737358(2)        & 0.2012504647824(3)        & 6.1267909226(3)       & 544.7845009(2)        & 545.1244240(5)\\
            & 0.201562974               & 0.201250465               & 6.12679095            & 544.785               & 545.124 \\
$(2,3)$     & 0.19718702654295(2)       & 0.19686565432911(8)       & 5.9972415652(1)       & 1178.74702791(2)      & 1179.4529688(1)\\
            & 0.197187028               & 0.196865656               & 5.99724165            & 1178.75               & 1179.45 \\
$(3,3)$     & 0.19310402152252(7)       & 0.192771518257(4)         & 5.876862410(2)        & 1979.084658(2)        & 1980.246050(8)\\
            & 0.193104023               & 0.192771520               & 5.87686248            & 1979.08               & 1980.25 \\
$(4,3)$     & 0.18929902980(2)          & 0.18895264869(6)          & 5.7652006889(6)       & 2884.65201(1)         & 2886.32533(6)\\
            & 0.189299034               & 0.188952656               & 5.76520094            & 2884.65               & 2886.33 \\
$(0,4)$     & 0.205528777980949(6)      & 0.20522321312906(3)       & 6.24365466610(2)      & 184.59247475(2)       & 184.7179604(1)\\
            & 0.205528778               & 0.205223214               & 6.24365467            & 184.592               & 184.718 \\
$(1,4)$     & 0.20088331251301(2)       & 0.2005706622275(5)        & 6.1057017310(3)       & 615.8801839(4)        & 616.262509(1)\\
            & 0.200883313               & 0.200570663               & 6.10570175            & 615.880               & 616.262 \\
$(2,4)$     & 0.1965457848848(2)        & 0.1962241242204(1)        & 5.97737426240(4)      & 1275.84569327(1)      & 1276.6080713(2)\\
            & 0.196545786               & 0.196224126               & 5.97737434            & 1275.85               & 1276.61 \\
$(3,4)$     & 0.192499569868(1)         & 0.192166602681(1)         & 5.8581666305(8)       & 2095.33663918(2)      & 2096.564775(2)\\
            & 0.192499571               & 0.192166604               & 5.85816669            & 2095.34               & 2096.56 \\
$(4,4)$     & 0.188729935323(7)         & 0.1883828767375(1)         & 5.747632304(5)        & 3013.980528(6)        & 3015.72739(3)\\
            & 0.188729937               & 0.188382879               & 5.74763241            & 3013.98               & 3015.73 \\
\hline\hline\end{tabular}
\end{table*}

\begin{table*}
\caption{Expectation values of
${R}_{dp}$, ${R}_{de}$, and ${R}_{pe}$ for HD$^+$ with $v=0-4$ and $L=0-4$. Korobov's results~\cite{Kor06} are listed in the second entry of each ro-vibrational state.}
\label{tb:hde:breit2}
\begin{tabular}{clllll}
\hline\hline
$(v,L)$ & $~~{R}_{dp}$ & $~~{R}_{de}$ & $~~{R}_{pe}$ \\\hline
$(0,0)$ & 5.35463051901711943(7)    & 1.174487825932605556(3)   & 1.170770145051139727(2) \\
        & 5.35463                   & 1.17449                   & 1.17077 \\
$(1,0)$ & 15.14482588936280077(6)   & 1.1504812637673131(2)     & 1.1433664050045(2) \\
        & 15.1448                   & 1.15048                   & 1.14337 \\
$(2,0)$ & 23.59899696623698(2)      & 1.1282362008927494(6)     & 1.1181235693136(2) \\
        & 23.5990                   & 1.12824                   & 1.11812 \\
$(3,0)$ & 30.817653008622(1)        & 1.10767720643593(3)       & 1.0949375427689(2) \\
        & 30.8177                   & 1.10768                   & 1.09494 \\
$(4,0)$ & 36.88807887511(1)         & 1.088739929649(8)         & 1.073718403917(3) \\
        & 36.8881                   & 1.08874                   & 1.07372 \\
$(0,1)$ & 5.589562255342723(1)      & 1.1735785001855910(1)     & 1.1697193799366195(4) \\
        & 5.58956                   & 1.17358                   & 1.16972 \\
$(1,1)$ & 15.350575279863064(1)     & 1.1496258585220391(3)     & 1.14238073961236855(7) \\
        & 15.3506                   & 1.14963                   & 1.14238 \\
$(2,1)$ & 23.777762089355404(4)     & 1.1274325102467439(1)     & 1.11720015102388(1) \\
        & 23.7778                   & 1.12743                   & 1.11720 \\
$(3,1)$ & 30.97134959374(1)         & 1.1069233223353(1)        & 1.09407389213715(9) \\
        & 30.9713                   & 1.10692                   & 1.09407 \\
$(4,1)$ & 37.0183653341(6)          & 1.08803423109(2)          & 1.07291239499(1) \\
        & 37.0184                   & 1.08803                   & 1.07291 \\
$(0,2)$ & 6.055104818249343733(4)   & 1.171769108405540(2)      & 1.16762928124835329(3) \\
        & 6.05510                   & 1.17177                   & 1.16763 \\
$(1,2)$ & 15.75802763752410(4)      & 1.147923888699637(7)      & 1.140420315240445(2) \\
        & 15.7580                   & 1.14792                   & 1.14042 \\
$(2,2)$ & 24.13150123361982(4)      & 1.12583357622946(1)       & 1.11536372709830(3) \\
        & 24.1315                   & 1.12583                   & 1.11536 \\
$(3,2)$ & 31.275189712792(3)        & 1.1054236322514(2)        & 1.0923565391832(2) \\
        & 31.2752                   & 1.10542                   & 1.09236 \\
$(4,2)$ & 37.275607599566(2)        & 1.086630566009(1)         & 1.07130988884988(7) \\
        & 37.2756                   & 1.08663                   & 1.07131 \\
$(0,3)$ & 6.74277630617217739(6)    & 1.16907791520748(2)       & 1.1645223841766(2) \\
        & 6.74278                   & 1.16908                   & 1.16452 \\
$(1,3)$ & 16.35924480624(3)         & 1.145392790044(2)         & 1.13750662824(1) \\
        & 16.3592                   & 1.14539                   & 1.13751 \\
$(2,3)$ & 24.6527816700128(9)       & 1.1234560568468(8)        & 1.112634817630(2) \\
        & 24.6528                   & 1.12346                   & 1.11263 \\
$(3,3)$ & 31.72221152212(6)         & 1.103194064780(2)         & 1.0898050864(1) \\
        & 31.7222                   & 1.10319                   & 1.08981 \\
$(4,3)$ & 37.65328339(4)            & 1.0845441798(5)           & 1.068929626(1) \\
        & 37.6533                   & 1.08454                   & 1.06893 \\
$(0,4)$ & 7.640242669140628(6)      & 1.16553169532956(5)       & 1.1604316913454(5) \\
        & 7.64024                   & 1.16553                   & 1.16043 \\
$(1,4)$ & 17.1426952788899(3)       & 1.1420581168039(4)        & 1.133671151272(3) \\
        & 17.1427                   & 1.14206                   & 1.13367 \\
$(2,4)$ & 25.3308170266259(7)       & 1.120324362118(2)         & 1.109043458742(1) \\
        & 25.3308                   & 1.120320                  & 1.10904 \\
$(3,4)$ & 32.30232200847(1)         & 1.100257957806(8)         & 1.086448221081(5) \\
        & 32.3023                   & 1.10026                   & 1.08645 \\
$(4,4)$ & 38.14194609(1)            & 1.0817974074(2)           & 1.0657990275(3) \\
        & 38.1419                   & 1.08180                   & 1.06580 \\
\hline\hline
\end{tabular}
\end{table*}

\begin{table*}
\caption{Numerical values of $Q({\bf r}_{1})$ for HD$^+$ with $v=0-4$ and $L=0-4$, where the second entry lists
Korobov's results~\cite{Kor06}.}
\label{tb:hde:qtm2}
\begin{tabular}{cllllll}
  \hline\hline
            & $~~v=0$ & $~~v=1$ & $~~v=2$ & $~~v=3$ & $~~v=4$ \\\hline
      $L=0$ & --0.1348622766081903(2) & --0.132113355129(1)    &--0.129551987571271(1) &--0.1271692904455(1) &--0.12495749557739(3) \\
            & --0.13486               & --0.13211              &--0.12955              &--0.12717            &--0.12496 \\
      $L=1$ & --0.134742001057392(5)  & --0.13199961143967(2)  &--0.1294445278685(4)   &--0.1270678977803(2) &--0.1248619827842(2) \\
            & --0.13474               & --0.13200              &--0.12944              &--0.12707            &--0.12486 \\
      $L=2$ & --0.13450284686085(8)   & --0.1317734666211(3)   &--0.1292308990657(3)   &--0.126866353373(3)  &--0.1246721506275(1) \\
            & --0.13450               & --0.13177              &--0.12923              &--0.12687            &--0.12467 \\
      $L=3$ & --0.1341475653900(1)    & --0.131437564457(6)    &--0.12891364214(1)     &--0.12656709998(4)   &--0.124390349(3)\\
            & --0.13415               & --0.13144              &--0.12891              &--0.12657            &--0.12439 \\
      $L=4$ & --0.1336801802328(6)    & --0.130995769843(4)    &--0.128496470584(6)    &--0.12617370653(3)   &--0.1240200070(4) \\
            & --0.13368               & --0.13100              &--0.12850              &--0.12617            &--0.12402 \\
  \hline\hline
\end{tabular}
\end{table*}

\begin{table*}
\caption{Numerical values of $Q({\bf r}_{12})$ for HD$^+$ with $v=0-4$ and $L=0-4$, where the second entry lists
Korobov's results~\cite{Kor06}.}
\label{tb:hde:qtm1}
\begin{tabular}{cllllll}
  \hline\hline
            & $~~v=0$ & $~~v=1$ & $~~v=2$ & $~~v=3$ & $~~v=4$ \\\hline
      $L=0$ &--0.1345911955685105(1) &--0.131838977797(3)    &--0.129272929986259(7) &--0.1268839283180(2) &--0.12466388687278(6) \\
            &--0.13459               &--0.13184              &--0.12927              &--0.12688            &--0.12466 \\
      $L=1$ &--0.13447097875759(1)   &--0.131725268104580(5) &--0.1291654760689(3)   &--0.1267825085601(1) &--0.1245683078029(8) \\
            &--0.13447               &--0.13172              &--0.12916              &--0.12678            &--0.12457 \\
      $L=2$ &--0.13423194023120(3)   &--0.1314991894516(1)   &--0.1289518569532(7)   &--0.126580907975(1)  &--0.124378340936(7) \\
            &--0.13423               &--0.13150              &--0.12895              &--0.12658            &--0.12438 \\
      $L=3$ &--0.133876827784(1)     &--0.13116338201(2)     &--0.12863460989(3)     &--0.1262815656(3)    &--0.12409633(1)\\
            &--0.13388               &--0.13116              &--0.12863              &--0.12628            &--0.12410 \\
      $L=4$ &--0.133409659744(2)     &--0.13072170536(5)     &--0.12821744286(1)     &--0.1258880438(2)    &--0.12372570179(5) \\
            &--0.13341               &--0.13072              &--0.12822              &--0.12589            &--0.12373 \\
  \hline\hline
\end{tabular}
\end{table*}

We summarize the contributions up to $R_{\infty}\alpha^5$ to two ro-vibrational transition frequencies
in Table~\ref{tb:hde.tot}, where the values of the Bethe logarithm are taken from Ref.~\cite{Kor12} and the $R_{\infty}\alpha^4$ order relativistic correction $E^{(4)}_{\rm{rel}}$ taken from Ref.~\cite{Kor08}. For the $(1,0)\rightarrow(0,0)$ transition,
the numerical uncertainty in $\Delta{E}_{\alpha^2}$ has been reduced from 1~kHz in Korobov's value to the present 8~Hz, which is
  due entirely to the uncertainties in the fundamental constants. The correction $\Delta{E}_{\alpha^3}$ to this transition has been obtained in Ref.~\cite{Kor12} and reproduced here. The recoil correction of $R_{\infty}\alpha^4(m/M)$ and higher
   contributes at the level of relative $10^{-10}\!-\!10^{-11}$, which causes a theoretical uncertainty of 1~kHz in $\Delta{E}_{\alpha^4}$. The largest uncertainty for the transition $(1,0)\rightarrow(0,0)$ comes from the theoretical uncertainty of $\Delta{E}_{\alpha^5}$. The uncertainty in $\Delta{E}_{\rm{nuc}}$ is due to the uncertainties in the proton and deuteron
   charge radii. For the transition $(4,3)\rightarrow(0,2)$, both experimental~\cite{Koe07} and theoretical~\cite{Kor08} results
   are available. In our calculation for this transition, although the uncertainties in $\Delta{E}_{\alpha^2}$ and $\Delta{E}_{\alpha^3}$ have been reduced to sub kHz, the total uncertainty in the transition frequency remains as large as 70~kHz, which is from the $\Delta{E}_{\alpha^5}$ term.

\begin{table*}
\caption{Summery of contributions to the HD$^+$ transition frequencies (in MHz).}
\label{tb:hde.tot}
\begin{tabular}{r@{\hspace{5mm}}r@{.}l@{\hspace{5mm}}r@{.}l@{\hspace{3mm}}r@{.}l}
\hline\hline
            & \multicolumn{2}{c}{$(4,3)\rightarrow(0,2)$}   & \multicolumn{4}{c}{$(1,0)\rightarrow(0,0)$}  \\\cline{4-7}
Author      & \multicolumn{2}{c}{Present} & \multicolumn{2}{c}{Present} & \multicolumn{2}{c}{Korobov \cite{Kor08}} \\
\hline
  $\Delta{E}_{\rm{nr}}$     &214\ 976\ 047&328\ 2(6)  &57\ 349\ 439&973\ 34(14)  &57\ 349\ 439&9717\\
 $\Delta{E}_{\alpha^2}$     &       3\ 411&702\ 93(4) &         958&276\ 694(8)  &         958&277(01)\footnotemark[1]\\
 $\Delta{E}_{\alpha^3}$     &        --891&610\ 9(1)  &       --242&126\ 26(4)   &       --242&125(02)\\
 $\Delta{E}_{\alpha^4}$     &          --6&457(1)     &         --1&748(1)       &         --1&748\\
 $\Delta{E}_{\alpha^5}$     &            0&388(74)    &           0&105(20)      &           0&105(19)\\
 $\Delta{E}_{\rm{nuc}}$     &          --0&462(6)     &         --0&125\ 2(17)   &         --0&125(2)\footnotemark[1]\\
\hline
 $\Delta{E}_{\rm{tot}}$     &214\ 978\ 560&88(7)      &57\ 350\ 154&355(22)      \\
$E_{\rm theor}$ \cite{Kor08}&214\ 978\ 560&88(7)      &\multicolumn{2}{c}{}      &57\ 350\ 154&355(21)\\
$E_{\rm expt}$ \cite{Koe07} &214\ 978\ 560&6(5)\\
\hline\hline
\end{tabular}
\footnotetext[1]{Obtained using the data of Table V in Ref. \cite{Kor06}.}
\end{table*}

In summary, we have presented an independent calculation of the Breit operators and the $Q$ terms for the
HD$^+$ ro-vibrational states ($v=0\!-\!4,L=0\!-\!4$), which provides a verification of previous theoretical results.

\begin{acknowledgments}
The authors would like to thank V.~I.~Korobov for helpful remarks. The work was supported by the NSFC under Grants No. 11004221 and No. 10974224, by the National Basic Research Program of China (973 Program) under Grant No. 2010CB832803, and by
NSERC of Canada. The work was carried out at the computing facilities of SHARCnet of Canada and Wuhan University of China.
\end{acknowledgments}



\end{document}